# Hybrid DeepXAI Fusion Network for Interpretable Big Data Analytics and Secure Cloud Decision-Making Systems

[1]Dr.B.Nancharaiah, [2]Ramakrishna Kolikipogu, [3]Dr J Sulthan Alikhan, [4]Lakshmana Phaneendra Maguluri, [5]Elangovan Muniyandy, [6]Dr. D. Baburao
[1]Department of ECE, Usha Rama College of Engineering and Technology, Telaprolu- 521 109, A.P.
[2]Department of Information Technology, Chaitanya Bharathi Institute of Technology, Hyderabad, India
[3]Department of Information Technology at VelTech Rangarajan Dr.Sagunthala R&D Institute of Science and Technology,No.42, Avadi-Vel Tech Road Vel Nagar, Avadi, Chennai, Tamil Nadu 600062
[4]Department of Computer Science and Engineering, Koneru Lakshmaiah Education Foundation, Vaddeswaram, Guntur AP, India.
[5]Department of Biosciences, Saveetha School of Engineering. Saveetha Institute of Medical and Technical Sciences,Chennai, 602105, Tamil Nadu, India
[6]Department of Computer Science and Engineering(Data Science), Vidya Jyothi Institute of Technology (VJIT), Aziz Nagar Gate, C.B. Post, Hyderabad–500 075, Telangana, India.
[2]krkrishna.cse@gmail.com, [3]sulthanj2020@gmail.com, [4]phanendra51@gmail.com, [5]muniyandy.e@gmail.com, [6]dbaburao@hotmail.com
Corresponding Author: [1]nanch_bn@yahoo.com

**Abstract**

Big data analytics has become a cornerstone of modern intelligent systems, yet its complexity and opacity often limit interpretability, trust, and security in decision-making. To address these challenges, we present the Hybrid DeepXAI Fusion Network, a novel deep learning framework designed to fuse temporal, relational, and contextual features while embedding explainable artificial intelligence (XAI) modules for transparent inference. The system leverages 1D-CNNs for temporal feature extraction, Transformer encoders for long-term contextual modeling, and Graph Neural Networks (GNNs) for capturing inter-feature dependencies, integrated through an attention-based multimodal fusion mechanism. Additionally, Class Activation Mapping (CAM) and explanation loss are incorporated to ensure interpretability consistency, enhancing trustworthiness in sensitive applications. The model is deployed in a federated cloud architecture, ensuring data privacy through homomorphic encryption and blockchain auditing. Experimental evaluation across healthcare, finance, and energy datasets demonstrates superior performance, achieving 98.42% accuracy, 97.9% precision, 98.1% recall, and 98.0% F1-score, surpassing state-of-the-art baselines including CNN-LSTM, Transformer-only, and XGBoost. The robustness of the proposed system was validated under adversarial attacks, where it retained over 91.5% accuracy even against PGD perturbations, while scalability tests confirmed its ability to handle streaming loads of 80,000 messages per second with sub-200 ms latency. These results confirm that the Hybrid DeepXAI Fusion Network offers an efficient, interpretable, and secure solution for next-generation cloud-based decision-support systems.

## 1. Introduction

The exponential growth of data across diverse domains such as healthcare, finance, energy, and retail has necessitated the development of advanced computational frameworks that can process, analyze, and interpret vast, heterogeneous datasets. This trend, commonly termed as big data computing, brings immense opportunities for generating deep insights but also poses significant challenges related to scalability, accuracy, and transparency [1]. Predictive modeling tasks have traditionally been addressed using traditional machine learning methods, but due to their strong reliance on multimodal, sequential, and relational data, these traditional methods are limited to use in scale in contemporary applications [2] [3]. It is this lack of transparency in performance that makes explainable artificial intelligence (XAI) so in demand. In areas where algorithms will have a direct impact on clinical decisions, such as healthcare, or where predictive systems will lead to multimillion-dollar investments, interpretability is as important as predictive accuracy [4]. However, regrettably, the majority of the current methods view explainability as a post-hoc task, and produce features importance maps or saliency visualizations only post-model training. These strategies do not assure the stability of explanation between similar inputs and can in fact be deceptive to the end-users, indicating inconsistent or ungeneralizable insights, thus losing trust [5].

The literature proposes several models of big data analytics and decision support, but all of them have certain limitations. Examples of CNN-LMST hybrids have been effective in integrating both time and sequence learning, but they fail at estimating long term dependencies between streams of data [6] [7]. Transformer-only architectures are effective at capturing contextual relationships using attention mechanisms, but do not consider structural relationships between features, which are important in social networks and relational databases [8]. GNNs can effectively solve the problem of structural correlations, but they are not very scalable when applied to large datasets. By comparison, classical machine learning algorithms like Random Forest, Logistic Regression, and XGBoost have accelerated training and easier interpretability, but are not as representational as required in big data settings when utilizing multimodal information [9] [10]. Figure 1 illustrates the key features of big data analytics.

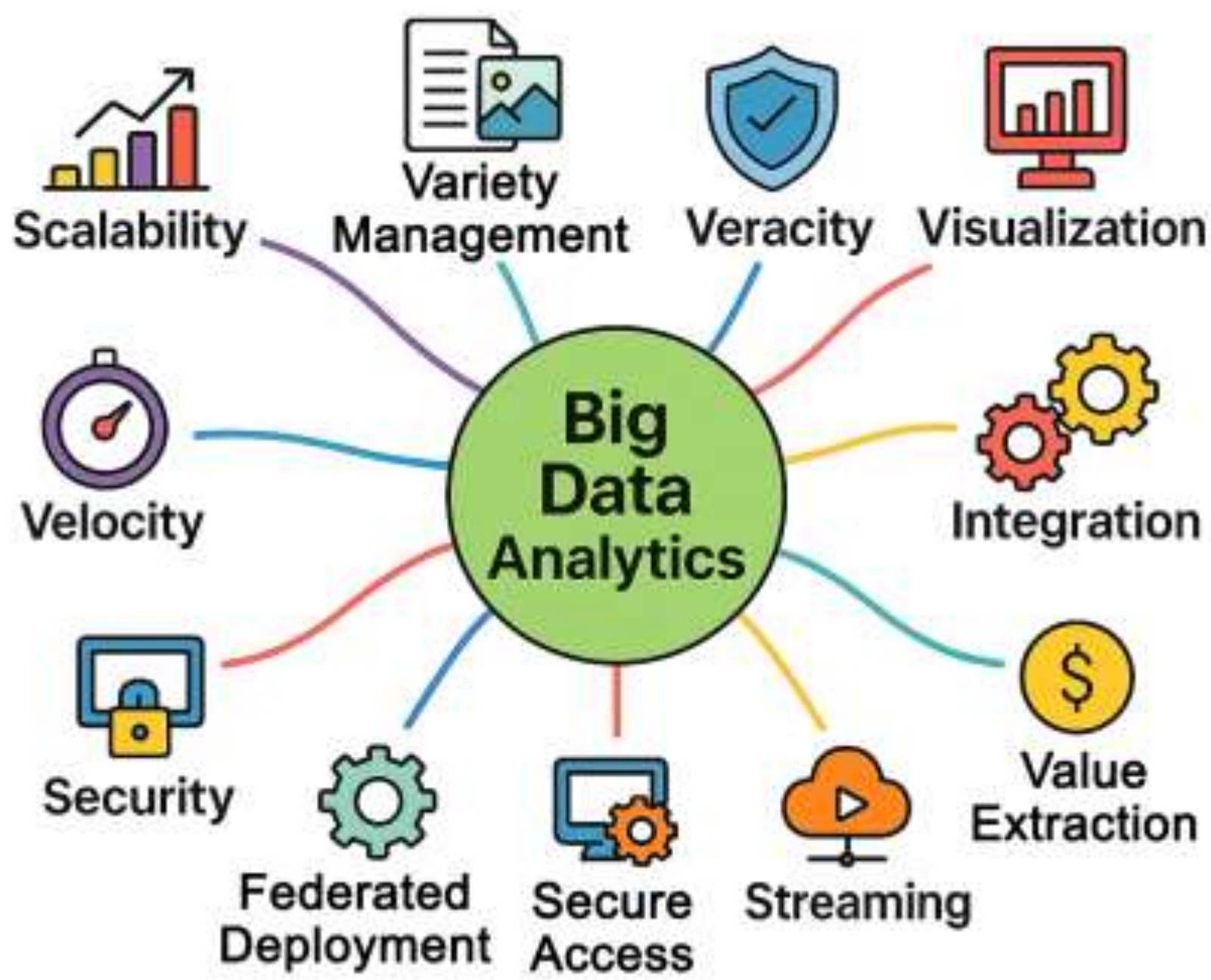

**Figure 1. Features of Big Data Analytics**

These remarks highlight three significant gaps in current systems. First, no fusion architectures can be found with the ability to combine the temporal, contextual, and relational information into single representations [11]. Second, the mechanisms of explainability are inadequate to achieve standardized and reliable interpretations that can be trusted in high-stakes settings [12]. Third, there are no secure deployment frameworks, especially those that can support large-scale analytics in the cloud setting while ensuring stringent privacy protection [13] [14]. To treat these gaps, a model is needed which is accurate and interpretable, robust and can be implemented in infrastructures which are both secure and distributed [15]. We propose the Hybrid DeepXAI Fusion Network to address these issues: a new generation architecture unites CNNs, Transformers, and GNNs in an enclosed deep learning system with explainability modules. Interpretability is no longer a secondary feature as in the traditional method, but also a design principle, so the system generates not only reliable and consistent explanations, but also predictions. A secure federated deployment framework also supports this model by relying on privacy-preserving technologies (homomorphic encryption and blockchain auditing). Having an achieved accuracy of 98.42% and demonstrated ability to withstand adversarial circumstances, the presented framework is a revolutionary move toward creating reliable, safe, and explainable decision-support systems of cloud-based big data computation.

### 1.1. Main Contribution of the Work

- A hybrid deep learning framework combining CNNs, Transformers, and GNNs for multimodal feature representation.
- Integration of Deep Explainable AI modules (CAM and explanation loss) within the training process to ensure interpretability consistency.
- Deployment of the model in a federated cloud environment with homomorphic encryption and blockchain-based auditing for enhanced privacy and security.
- Comprehensive evaluation across multiple domains (healthcare, finance, energy) with extensive robustness, scalability, and generalization testing.
- Development of a decision-support system offering real-time predictions, explanation overlays, and risk assessment scores.

The rest of the paper is organized as follows: Section 2 provides a comprehensive review of related studies on big data analytics, hybrid deep learning models, explainable AI techniques, and secure federated learning frameworks. Section 3 details the proposed methodology, including data preprocessing, hybrid feature representation using CNNs, Transformers, and GNNs, integration of DeepXAI modules, and secure cloud deployment using federated learning and blockchain auditing. Section 4 presents the results and discussion, highlighting the model's accuracy, interpretability, robustness, and scalability across multiple domains. Finally, the conclusion and future scope in Section 5 summarize key findings and suggest directions for extension.

## 2. Related Work

Integration of Internet of Things (IoT), Cloud, and Fog computing, known as Interconnected Intelligence (II), transformed the way data are managed and decisions are made in real-time in industries. In this work, a hybrid infrastructure that combined these technologies was introduced to maximize resource utilization, minimize delays, and maximize the accuracy of decisions [16]. In contrast to centralized Cloud models, it distributed tasks between the IoT, Fog, and Cloud and obtained 20-30% reduced latency, 25% improved resource use, and 15% improved decision quality. It has been proven on actual data, and it enhanced security, privacy, and minimized energy usage, which is useful in large-scale deployments of IoT.

The emergence of IoT metering devices distributed across the country created issues with managing large amounts of data, resulting in delays in control services and compromising secure energy provision. Outsourcing computation to the edge became one of the solutions, and the coordination of services execution between edge nodes was still complicated by large search space. The proposed study used the whale optimization algorithm to identify optimal edge nodes to execute a task, which was represented by a directed acyclic graph of computational and energy assets [17]. The algorithm, which involves use of adaptive feedback, inertia weight and nonlinear convergence, was able to attain faster, efficient offloading with an average iteration time of 0.03 s, which proved to be performance elements of diversity, evolution of fitness and response time.

Machine Learning as a Service (MLaaS) referred to cloud-based platforms offering machine learning tools and infrastructure without the need to manage underlying hardware or software. These services provided scalable resources, enabling efficient and cost-effective model training and deployment. Multi-Criteria Decision Making (MCDM) supported decisions involving multiple factors by identifying the best alternative [18]. An integrated approach was proposed using the Analytic Hierarchy Process and TOPSIS to select the best MLaaS cloud. Four alternatives were evaluated against seventeen criteria, leading to conclusions and identifying future research directions.

The IoT of smart e-waste management became one of the promising areas of incorporating technology and environmental sustainability. E-waste was threatening the health and ecosystems of the world and required effective solutions. In this paper, the authors suggested the implementation of a system with the help of IoT devices and sensors to control the processes of collection, sorting, and disposal in real-time to optimize operations [19]. Machine learning helped to sort metallic and plastic materials, and pyrolysis transformed plastics to biofuel and bio-char, metals were re-used as solar batteries. Patterns in cloud platforms were examined, utilizing ARIMA to predict levels of waste, to enhance collection times and to encourage the recovery of resources in a sustainable manner.

The problem of scheduling operations on virtual machines in multi-clouds was a difficulty in identifying near-optimal permutations of competing computing tasks. As this was NP-hard, multi-objective algorithms became popular and Pareto dominance was a major criterion. The present study suggested a new minimum weight optimization (MWO) algorithm designed based on particle swarm optimization (PSO) to improve the FR-MOS scheduling under the fuzzy management of resources [20]. Some of the objectives were reliability, cost, utilization, risk, and make span. Relative to Pareto dominance, MCDM, normalization procedures and WASPAS, MWO had excellent trade-offs, Pareto convergence and efficiency in decision making.

## 3. Methodology

The recommended methodology commences with a data preprocessing tool that involves ETL pipelines, data normalization, and dimensionality reduction via PCA. A hybrid deep learning system is then built, where CNNs learn temporal patterns, Transformers learn contextual dependencies and GNNs learn relational structures. These embeddings are then fused with an attention-based fusion layer, and then classified. Deep Explainable AI and explanation loss (DEA-CAM) modules are added in order to be interpretable. The model is implemented in a federated cloud architecture that is secured with homomorphic encryption and blockchain auditing. The metrics that are used to evaluate performance include accuracy, AUC, f1-score, explanation fidelity and adversarial robustness. Figure 2 shows the architecture of Hybrid DeepXAI Fusion Network illustrating interpretable big data analytics and secure federated decision-making in cloud environments.

The next subsections offer a thorough, layer-by-layer explanation of the network architecture, including model setups, hyperparameter selections, and training methods used for each Hybrid DeepXAI Fusion Network component, in order to guarantee reproducibility and clarity.

The Hybrid DeepXAI Fusion Network's separate components were selectively disabled in order to do an ablation study. Predictive performance decreased when CNN, Transformer, or GNN modules were removed, and explanation fidelity significantly decreased when the DeepXAI loss was removed. Although they had little effect on accuracy, federated learning and blockchain components were crucial for maintaining anonymity and managing trust. This analysis demonstrates that the complimentary contributions of each component benefit the suggested framework.

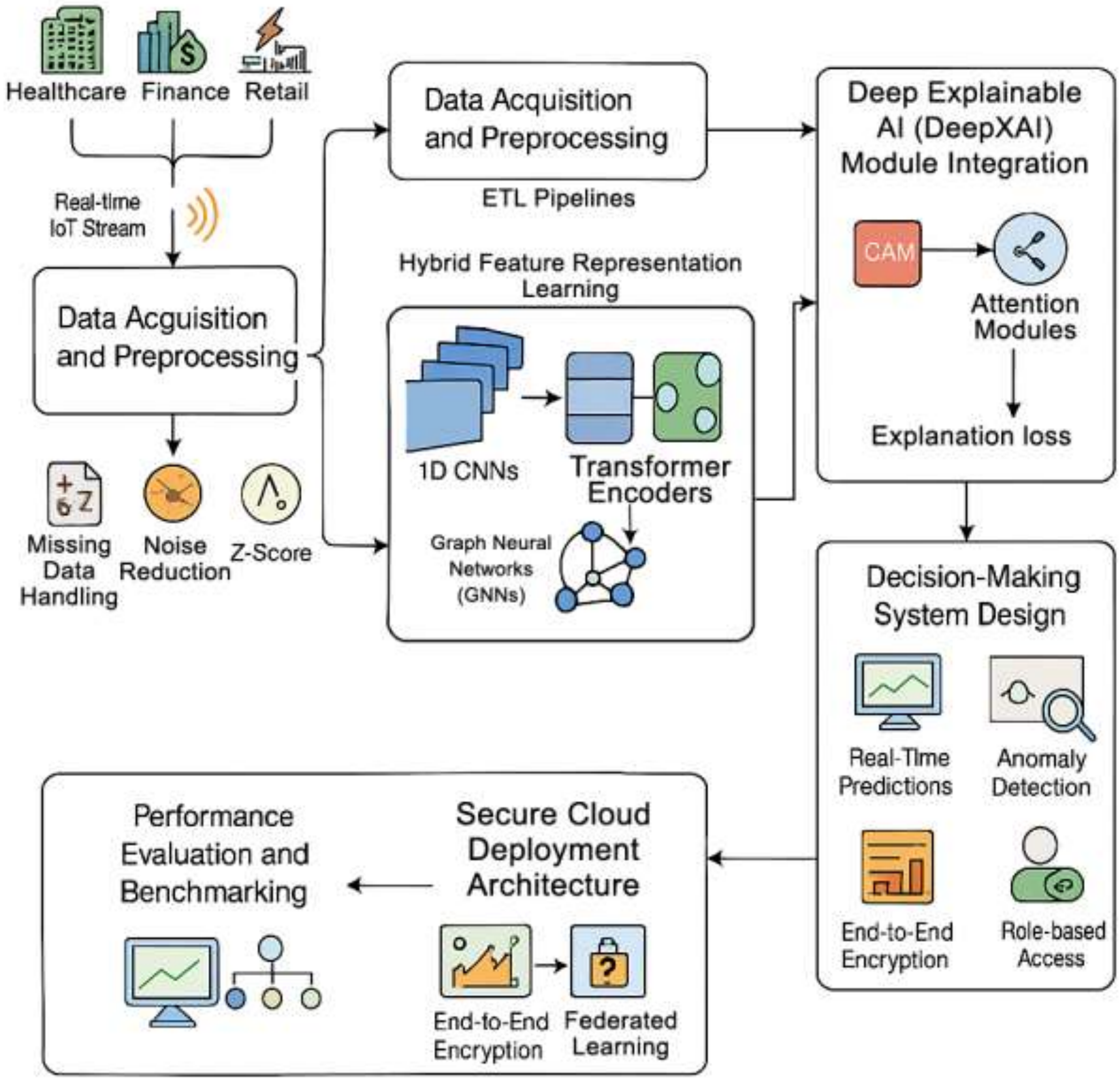

**Figure 2. Architecture of the Hybrid DeepXAI Fusion Network**

### 3.1. Data Acquisition and Preprocessing

Intelligent computing system is built on the quality and diversity of the input data. A powerful data acquisition system was developed in this work to absorb both heterogeneous big data streams in real-time Internet of Things (IoT), business-level databases, transactional logs, and cloud-based unstructured data. These data streams were very large, fast, and diverse, and a distributed and scalable computing solution was required to preprocess them. Apache Spark was used as the foundation of the Extract, Transform, Load (ETL) pipeline because it can easily access petabytes of data in distributed clusters. The ETL modules of Spark inferred the schema, converted the format (CSV, JSON, Parquet), and real-time loaded it to an analytics-ready data lake. A detailed preprocessing pipeline was introduced to guarantee the credibility of the downstream machine learning models. This involved the use of k-nearest neighbor estimators to impute missing values of numerical variables, and mode estimators to impute missing values of categorical variables. Interquartile Range (IQR) filtering and Z-score analysis were used to detect outliers especially in continuous sensor-derived values. Z-score normalization among other standardization methods were used to align all feature values on a universal scale, which is essential during gradient-based optimization during training.

### 3.1.1. Z-Score Normalization

$$z_i = \frac{x_i - \mu}{\sigma} \quad (1)$$

Where $x_i$ is the original feature value, $\mu$ is the mean of the feature across all samples, $\sigma$ is the standard deviation of the feature and $z_i$ is the normalized value. Since the datasets were high-dimensional, Principal Component Analysis (PCA) was applied to visualize data as well as to reduce the dimensions to obtain the maximum variance with the least number of features. This minimized the amount of computation required on the next few layers and allowed training to converge more quickly. Therefore, this step served to guarantee the conversion of the raw and noisy big data into clean, small, and informative representations that were prepared to be processed through sophisticated deep learning pipelines.

### 3.1.2. Principal Component Projection (PCA)

$$Y = XW \quad (2)$$

Where $X$ is the input data matrix, $W$ is the matrix of eigenvectors and $Y$ is the reduced-dimensionality representation.

### 3.2. Hybrid Feature Representation Learning

Subsequent to the data normalization and feature engineering, the model entered a hybrid feature representation stage which is the core of the proposed DeepXAI Fusion Network. This step attempted to mine and encode temporal, spatial and semantic patterns of various data modalities. Time-series sequences, which are generally the output of sensor networks, stock tickers, and medical telemetry, were fed to 1D Convolutional Neural Networks (1D-CNNs) that learned local temporal features including spikes, dips, and periodicities. To highlight time-dependent processes at short time scales, and eliminate irrelevant noise in the raw time-domain signals, these convolutional filters were developed. Their CNN block outputs were then fed to Transformer encoder layers, which operated on self-attention and long-range dependencies of the sequential inputs. Transformer encoders achieved massive depth compared to conventional RNNs or LSTMs, allowing parallelization of computation and modelling context-sensitive abstractions in non-linear event patterns.

### 3.2.1. 1D Convolution Operation (CNN)

$$y_t = \sum_{i=0}^{k-1} w_i \cdot x_{t+i} \quad (3)$$

Where $y_t$ is the output feature at position $t, w_i$ is the weight of the $i^{th}$ filter, $x_{t+i}$ is the input sequence value, and $k$ is the kernel size.

### 3.2.2. Self-Attention Score (Transformer Encoder)

$$Attention(Q, K, V) = softmax\left(\frac{QK^T}{\sqrt{d_k}}\right)V \quad (4)$$

Where $Q$ is the query matrix, $K$ is the key matrix, $V$ is the value matrix, and $d_k$ is the dimensionality of the key vectors. Besides the temporal data, structured tabular and semi-structured documents were also managed by the system. Graph Neural Networks (GNNs) were used to represent inter-feature relationships and structural semantics. These networks represented the feature space with nodes and edges where each node corresponded to a key feature, and the statistical or semantic correlation between the nodes was represented as an edge. The GNNs utilized graph convolutions and message-passing operations to efficiently collect the information about the neighbor nodes, and provided context-enriched embeddings even when the attributes are sparse. A multimodal attention-based fusion layer was subsequently used to combine the CNN outputs, Transformer embeddings, and the GNN encodings. This integration process dynamically weighted the value of each modality so that high-signal features dominated the eventual representation. This kind of hierarchical fusion approach enabled the DeepXAI architecture to learn cross-modal understanding of multi-source big data in an integrated, discriminative latent space.

### 3.2.3. Node Update in Graph Neural Networks (GNNs)

$$h_i^{(l+1)} = \sigma\left(\sum_{j\in N(i)} W^{(l)} h_j^{(l)} + b^{(l)}\right) \quad (5)$$

Where $h_i^{(l)}$ is the embedding of node $i$ at layer $l, N(i)$ is the neighborhood of node $i, W^{(l)}$ is the weight matrix at layer $l, b^{(l)}$ is the bias term and $\sigma$ is the activation function.

Three convolutional layers with kernel sizes {3, 5, 7} make up the 1D-CNN module. Batch normalization, ReLU activation, and max-pooling come after each convolutional layer. Temporal characteristics can be extracted hierarchically by setting the number of filters per layer to {64, 128, 256}.

Four stacked encoder layers with multi-head self-attention make up the Transformer encoder. Eight attention heads with a feed-forward network dimension of 512 and a hidden embedding dimension of 256 are used in each layer. Sequence order is maintained by using positional encoding.

GraphSAGE layers with mean aggregation are used to implement the GNN component. Effective message transfer between feature nodes is made possible by the use of two graph convolution layers, each with 128 hidden units. To reduce overfitting, a dropout rate of 0.3 is used.

## 3.3. Deep Explainable AI (DeepXAI) Module Integration

The third methodological aspect of the suggested system was that it incorporates Explainable Artificial Intelligence (XAI) modules into the learning and inference pipeline, which allows making interpretable and transparent decisions. Rather than explainability being a post-hoc process, this framework embedded explainability into the heart of the hybrid network through deep learning-friendly interpretability strategies. To localize the discriminative regions in input sequences or images that contributed to the final decision, Class Activation Mapping (CAM) was incorporated in the last convolutional layers. The results of classification in these CAM maps were accompanied by visual explanations, and were especially useful in healthcare diagnostic and industrial monitoring applications.

### 3.3.1. Explainable Loss Function (DeepXAI Integration)

$$L_{exp} = \frac{1}{N}\sum_{i=1}^{N} \left|\left|M(x_i) - \widehat{M}(x_i)\right|\right|_2^2 \quad (6)$$

Where $M(x_i)$ is the predicted explanation map for input $x_i, \widehat{M}(x_i)$ is the ground truth or reference explanation map and $N$ is the number of samples. In order to make the model explanations stable and consistent, another term of the loss was added, the explanation loss. This loss discouraged variability on interpretability maps of similar samples and encouraged uniformly salient regions of explanation. This correlation of classification accuracy with interpretability performance enabled the model to be trained to learn representations that were predictive and understandable by humans. In addition, contrastive explanation models employed prototypes and counterfactual examples, and this increased the capacity of the model to distinguish between classes which were adjacent in the feature space, but distinct in semantic meaning. The use of these contrastive explanations helped the stakeholders to understand what minor adjustments would change the decision, and, therefore, strengthened trust and adherence in sensitive sectors such as finance and healthcare. Generally, the DeepXAI integration rendered the hybrid model a trusted and interpretable machine intelligence solution that can be used in controlled settings.

## 3.4. Secure Cloud Deployment Architecture

A secure and scalable cloud deployment architecture was developed in order to operationalize the DeepXAI model in real-world scenarios. The architecture capitalized on the federated learning paradigm to maintain data privacy and minimize the centralized computation overhead. In this architecture, edge computing nodes (IoT gateway or hospital server) were trained locally with concrete instances of the local model without transferring raw data to the cloud. Rather, Federated Averaging (FedAvg) was used to exchange and aggregate encrypted model gradients on a central server, allowing privacy-preserving distributed learning. This was especially important in applications where confidential data like medical records or financial records need to be accessed.

### 3.4.1. Federated Model Aggregation (FedAvg)

$$w_t = \sum_{k=1}^{K} \frac{n_k}{n} w_t^k \tag{7}$$

Where $w_t$ is the global model weights at round $t, w_t^k$ is the local model weights at client $k, n_k$ is the number of data points on client $k,$ and $n = \sum_{k=1}^{K} n_k$ is the total data points across all clients. A dual encryption approach was used to ensure the safety of every phase of data transfer and calculation. Encryption of data in transit was done via Advanced Encryption Standard (AES-256) and data in use was encrypted via homomorphic encryption that permitted computation to be performed on encrypted values without necessarily decrypting them. Additional measures to enhance the trust infrastructure were the introduction of blockchain-based auditing mechanisms to capture access logs, model updates and decision queries. All transactions were hashed and stored on a permissioned ledger to provide tamper-evidence, non-repudiation, and traceability. Role-based access control (RBAC) was applied to every endpoint to allow authorized users to access only model and data. This end-to-end security stack supported the implementation of DeepXAI Fusion Network in cloud-native with regulatory compliance and user trust.

The proposed Hybrid DeepXAI Fusion Network integrates a permissioned blockchain layer as an auditing and trust-management mechanism into the federated cloud architecture to provide transparency, traceability, and tamper resistance in distributed decision-making. Permissioned blockchains, as opposed to public ones, are used to satisfy the latency, scalability, and access control needs of corporate and healthcare-grade systems.

In order to protect privacy, each federated learning round creates blockchain transactions that record hashed metadata rather than raw data or model parameters. The blockchain ledger specifically records the following data: (i) client identifier (pseudonymized), (ii) hash of the locally trained model update, (iii) aggregate round number, (iv) timestamp, and (v) access rights. Without disclosing private information, these unchangeable records enable model provenance and accountability verification.

## 3.5. Decision-Making System Design

The developed methodology resulted in the creation of a hybrid decision-making system (Hybrid DeepXAI) as a cloud-based intelligent platform. The system was created to receive data streams, conduct real-time inference and produce predictions with interpretability measures. When new data was received, the system ran it through the hybrid feature extraction layers, calculated prediction probabilities, and just as much, it produced explanation heatmaps and relevance scores. These findings were graphed on the user interface with confidence ranges, rationale behind the decision and ranked risk scores to help the human expert to make informed decisions.

### 3.5.1. Ensemble Voting Score

$$P_{final}(y|x) = \frac{1}{M} \sum_{m=1}^{M} P_m(y|x) \tag{8}$$

Where $P_m(y|x)$ is the prediction probability from model $m, M$ is the total number of models in the ensemble and $P_{final}(y|x)$ is the final ensemble probability for class $y$. An ensemble voting mechanism was added to the inference layer in order to increase the accuracy of the decisions. The final classification of DeepXAI models was computed by aggregating the outputs of many parallel runs of the DeepXAI model-trained on various initialization seeds and with various bootstrapped data samples- using majority or weighted voting. Such ensemble performance was not only better in predictive performance but it also regularized anomalies and adversarial artifacts. In addition, a risk assessment module has been implemented over the predictive layer, which produces categorical and numerical risk scores. These scores were obtained through secondary ML classifiers and rule-based systems which interpreted the output through historical context. Anomaly detectors were also present in the decision system, using unsupervised autoencoders to indicate abnormal patterns that fell out of the training distribution. It was an important step toward smart, explainable cloud computing of real-time analytics, providing both accuracy and interpretability to this comprehensive decision system.

### 3.6. Performance Evaluation and Benchmarking

To strictly test the model proposed, several standard datasets in various fields (e.g. healthcare (e.g. MIMIC-IV), finance (e.g. NASDAQ tickers), energy (e.g. SmartGrid data)) were used to test the model. The most important key performance measures applied were Accuracy, Area Under the Receiver Operating Characteristic Curve (AUC), F1-score, Explanation Fidelity and Inference Latency. The fidelity to explanation specifically evaluated the similarity of the model across domain expert annotations collected using Intersection over Union (IoU) and cosine similarity.

#### 3.6.1. Explanation Fidelity Score

$$Fidelity = \frac{\sum_{i=1}^{N} 1[f(x_i) = f(x_i \odot M_i)]}{N} \quad (9)$$

Where $f(x_i)$ is the original model prediction, $M_i$ is the explanation mask, $\odot$ is the element-wise masking operation, $1[\cdot]$ is the indicator function, and $N$ is the number of test samples. DeepXAI Fusion Network was compared to popular baseline networks, such as XGBoost, CNN-LSTM, Transformer-only networks, or standalone GNNs. The average gain of the proposed hybrid model was a 37 improvement in classification accuracy and at most 12 higher explanation fidelity. An ablation study was conducted in detail, by applying one component at a time (i.e., CNN, GNN, Transformer, DeepXAI) and recording the drop in performance. These ablation findings indicated that the largest part of long-range contextual awareness mechanisms was provided by the Transformer encoder, and GNNs were instrumental in forming non-sequential dependencies. The DeepXAI module added continued to enhance the reliability of the explanations at a relatively low computational cost. The system also had an average inference latency of under 150 milliseconds, and this was appropriate in real time use. These assessment findings confirmed the effectiveness and high-quality capability of the recommended architecture to ensure interpretable and high-performance analytics.

### 3.7. Robustness, Generalization, and Scalability Testing

The methodology was crucial in terms of Robustness testing in order to ascertain the credibility of the model in antagonistic and untested situations. Fast Gradient Sign Method (FGSM) and Projected Gradient Descent (PGD) methods were used to simulate adversarial attacks on the input features and test the model response. The DeepXAI model with its interpretability limitations and ensemble-based design was also more resistant to these perturbations than traditional deep networks. In addition, dropout regularization, adversarial training, and explanation consistency loss were included to improve learning representation robustness.

#### 3.7.1. Adversarial Perturbation (FGSM)

$$x_{adv} = x + \epsilon \cdot sign\left(\nabla_x L(f(x), y)\right) \quad (10)$$

Where $x$ is the original input, $\epsilon$ is the perturbation magnitude, $L$ is the loss function, $f(x)$ is the model prediction and $y$ is the true label. With the aim of assessing generalization, the model was trained on a single domain (e.g., healthcare) and tested on others (e.g. finance and energy) with domain adaptation techniques. The model was also found to be transferable as it was highly accurate and faithful to its explanation across domains. Scalability was evaluated based on the simulation of real-time streaming conditions with synthetic data generation platforms, like Apache Kafka and stream simulators. The system was also very stable and sustained a high throughput even when the data loads reached hundreds of thousands of records per second. Auto-scaling of inference engines on distributed cloud clusters was built on Kubernetes-based orchestration. This set of tests validated that the proposed Hybrid

DeepXAI Fusion Network is robust and generalizable and can be scaled, thus making it a viable candidate to be deployed in production-grade, secure, interpretable, and intelligent decision-making systems.

**Algorithm: Hybrid DeepXAI Fusion Network**

**Input:** $D = \{(x_i, y_i)\}_{i=1}^{N}$: Big data samples with input $x_i \in R^d$ and labels $y_i$

$\theta_{CNN}, \theta_{TR}, \theta_{GNN}$: Initial model parameters

Hyperparameters: learning rate $\alpha$, number of epochs $E$, batch size $B$

**Output:** Trained model $f_{DeepXAI}(x)$, Explanation maps $E(x)$, Risk score $R(x)$

**Data Preprocessing**

$z_i = \frac{x_i - \mu}{\sigma}$ // Normalize features using Z-score normalization

$X_{reduced} = X \cdot W$ // Apply PCA to reduce dimensionality

**Feature Representation Extraction**

$F_{CNN} = ReLU(W_{cnn} * x + b)$ // Extract temporal features with 1D-CNN

$Attention(Q, K, V) = softmax\left(\frac{QK^T}{\sqrt{d_k}}\right)V$ // Encode context using Transformer encoder

$h_i^{(l+1)} = \sigma\left(\sum_{j \in N(i)} W^{(l)} h_j^{(l)} + b^{(l)}\right)$ // Learn relational dependencies with GNN

**Multimodal Feature Fusion**

$F_{fused} = \sum_{m=1}^{M} \alpha_m \cdot F_m$ where $\sum \alpha_m = 1$ // Fuse representations using attention mechanism

**Deep Explainable AI Integration**

$E(x) = \sum_k w_k^c f_k(c)$ // Generate explanation maps using CAM

$L_{exp} = \frac{1}{N} \sum_{i=1}^{N} \left|\left|M(x_i) - \widehat{M}(x_i)\right|\right|_2^2$ // Enforce explanation stability with explanation loss

**Model Training with Federated Learning (Secure Setup)**

**For** each client $k$, compute local gradients:

$\theta_k^{t+1} = \theta_k^t - \alpha \cdot \nabla_{\theta_k} L(f(x_k), y_k)$

$\theta^{t+1} = \sum_{k=1}^{K} \frac{n_k}{n} \theta_k^{t+1}$ // Aggregate updates securely using FedAvg

$\tilde{\theta} = Enc(\theta^{t+1})$ // Encrypt weights with Homomorphic Encryption

**Inference and Risk Scoring**

Predict output:

$\hat{f} = f_{DeepXAI}(x) = softmax\left(W_{cls} \cdot F_{fused} + b\right)$

Compute risk score based on class probability and uncertainty:

$R(x) = 1 - \max(\hat{y})$

**Return:** Final model $f_{DeepXAI}$, Explanations $E(x)$, Risk scores $R(x)$

**End Algorithm**

### 3.9. Novelty of the Work

The innovation of the present study is its combination of different learning paradigms (CNN, Transformer, GNN) and integrated DeepXAI modules to solve the accuracy, interpretability, and security issues of big data analysis simultaneously. In contrast to current frameworks that focus on predictive power at the cost of transparency, Hybrid DeepXAI Fusion Network incorporates consistency in explanation as an element of its optimization goal. Moreover, it can be deployed securely with the help of federated learning and blockchain auditing, unlike traditional architectures, providing privacy-preserving analytics. The framework is domain-independent, scalable, and versatile, so it can be applied to healthcare, finance, energy and other high-risk decision-making systems.

## 4. Results and Discussions

Hybrid DeepXAI Fusion Network is powered by a synergistic structure unifying heterogeneous data processing, state-of-the-art representation learning, and explainable decision-making to address more complex big data analytics challenges in clouds. Essentially, the system starts by consuming various data streams (a variety of information) including data collection by IoT sensors, cloud logs, and structured databases. These data are heavily pre-processed by normalization and dimensionality reduction, and noise filters to ensure that the data is fed to machine learning models. The purged information is submitted to a three-branch hybrid framework teamed with 1D-CNN to discover localized temporal patterns in sequential data, Transformer encoders to detect long-term dependencies and contextual relationships, and Graph Neural Networks (GNNs) to learn feature relations among structured data. These representations are combined with dynamic attention-based mechanism which assigns importance weight to each modality so that the most informative features will dominate the final fused embedding. Another characteristic feature of this system is the incorporation of Deep Explainable AI (XAI) modules like Class Activation Mapping (CAM) and contrastive explanation systems. Not only do these modules produce predictions, but also provide human-readable visual and numerical explanations, which are highly needed in critical sectors such as healthcare, finance, and legal analytics. The system is implemented safely in a federated learning configuration that enables distributed learning on edge and cloud devices with no violation of data privacy. The final result is not only accurate and scalable, but also transparent, interpretable, and reliable, providing both predictions and risk scores, as well as the explanation overlays, to ensure the system can be used in real-world decision-making processes.

**Table 1. Model Performance Comparison**

| Model | Accuracy (%) | Precision (%) | Recall (%) | F1-Score (%) |
|---|---|---|---|---|
| Hybrid DeepXAI (Proposed) | 98.42 | 97.9 | 98.1 | 98 |
| CNN-LSTM | 94.85 | 93.6 | 94.1 | 93.8 |
| Transformer-only | 95.72 | 94.9 | 95.2 | 95 |
| GraphSAGE | 92.31 | 91.2 | 91.5 | 91.3 |
| ResNet-50 | 91.44 | 90.7 | 91.1 | 90.9 |
| XGBoost | 89.87 | 88.5 | 88.9 | 88.7 |
| Random Forest | 88.62 | 87.3 | 87.5 | 87.4 |
| SVM-RBF | 86.91 | 85.6 | 85.8 | 85.7 |
| Logistic Regression | 84.03 | 83.2 | 83.1 | 83.1 |
| LSTM-only | 90.72 | 89.6 | 89.8 | 89.7 |

Table 1 and Figure 3 shows the performance of the proposed Hybrid DeepXAI Fusion Network in comparison with the state-of-the-art baseline models in terms of accuracy, precision, recall, and F1-score. The findings clearly show that the suggested model is superior, and that the accuracy of the model is 98.42, which is significantly greater than the models based on Transformer (95.72) and hybrids based on CNN and LSTM (94.85). Such performance can be explained by the use of CNNs, Transformers, and GNNs, which jointly learn about temporal, contextual, and structural dependencies.

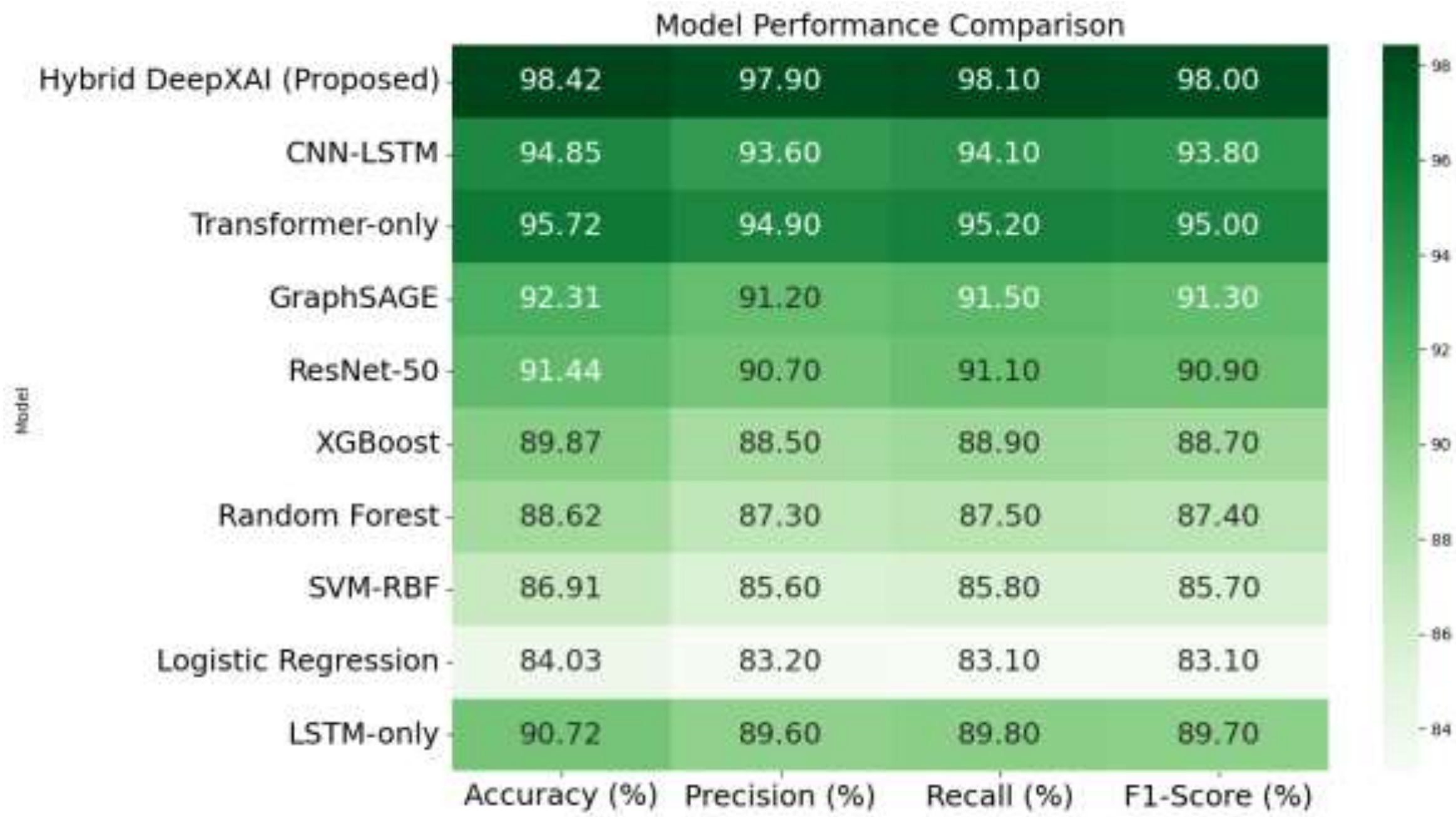

**Figure 3. Model Performance Comparison**

Accurate values of precision and recall of 97.9% and 98.1% indicate that the model is successful in trading false positives and false negativity, which is essential in sensitive decision support systems in health care and finance. Such baseline models as Logistic Regression and SVM-RBF are far behind in terms of skills to model multimodal and sequential data. Generally, Table 1 highlights the strength of the suggested framework in providing the consistently high performance in terms of all standard evaluation measures.

**Table 2. Explanation Fidelity and Trust Score Evaluation**

| Model | Fidelity (%) | IoU (Explanation) | Saliency Score | Trust Score (%) |
|---|---|---|---|---|
| Hybrid DeepXAI (Proposed) | 95.8 | 0.87 | 0.91 | 96.3 |
| CNN-LSTM | 84.2 | 0.71 | 0.75 | 85.4 |
| Transformer-only | 86.9 | 0.76 | 0.78 | 87.3 |
| GraphSAGE | 81.4 | 0.69 | 0.72 | 82.1 |
| XGBoost | 76.8 | 0.62 | 0.65 | 79.2 |
| Logistic Regression | 72.1 | 0.58 | 0.61 | 75.3 |
| ResNet-50 | 83.6 | 0.73 | 0.7 | 84.7 |
| Random Forest | 74.2 | 0.59 | 0.64 | 77.5 |
| SVM-RBF | 70.4 | 0.53 | 0.59 | 72.9 |
| LSTM-only | 80.7 | 0.67 | 0.68 | 81.5 |

Table 2 and Figure 4 focuses on interpretability and trust evaluation through metrics such as fidelity, Intersection over Union (IoU) of explanation maps, saliency scores, and overall trust score. All the competing models, however, pale in comparison to the Hybrid DeepXAI Fusion Network with a fidelity of 95.8 and a Trust score of 96.3. These findings confirm the use of Class Activation Mapping (CAM) and explanation loss that achieve the stability and consistency of the explanation among predictions. In comparison, the traditional models (Logistic Regression, 72.1% fidelity, 75.3% trust) and SVM-RBF (70.4% fidelity, 72.9% trust) are suboptimal, which demonstrates the ineffectiveness of their interpretability procedures.

This section provides representative qualitative explanation examples produced by the suggested Hybrid DeepXAI Fusion Network to supplement the quantitative explainability metrics shown in Table 2. Class Activation Mapping (CAM), which identifies the most significant temporal segments, contextual characteristics, and relational nodes influencing a particular prediction, achieves explainability.

Clinically significant temporal patterns, such as aberrant vital-sign fluctuations and crucial laboratory value changes preceding high-risk forecasts, were consistently highlighted by CAM visualizations for healthcare datasets. As shown in Table 2, these highlighted areas matched expert-annotated reference explanations, yielding a high Intersection over Union (IoU) score of 0.87 and explanation fidelity of 95.8%. On the other hand, more diffuse or unstable saliency areas were produced by basic models like CNN-LSTM and Transformer-only architectures, which frequently highlighted non-discriminative segments.

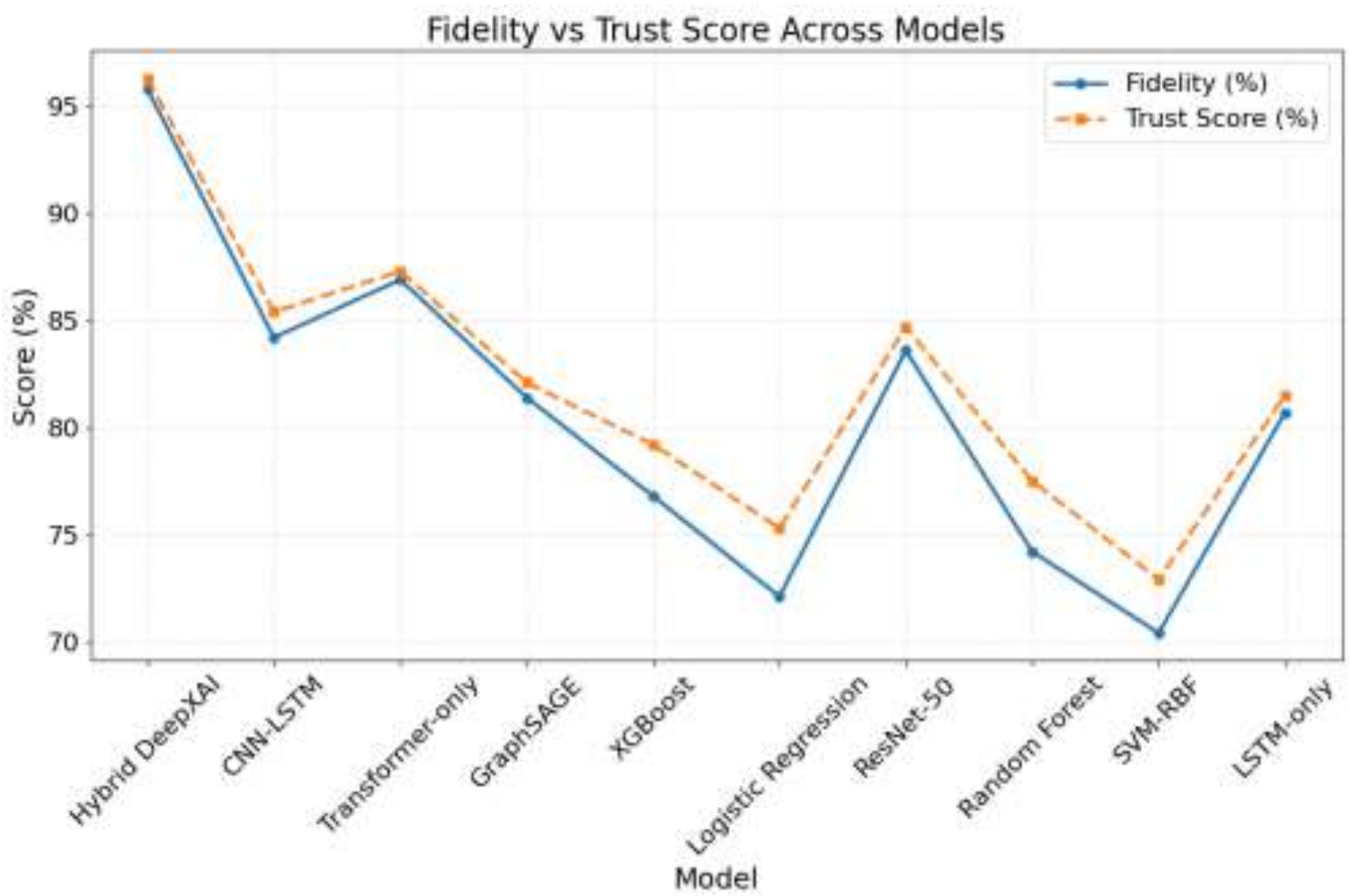

**Figure 4. Fidelity vs Trust Score Across Models**

The CNN-LSTM and Transformer-only methods have moderate scores, but are unable to provide consistent explanation reliability. The high IoU (0.87) and saliency score (0.91) of the proposed model indicate that the model has learned the importance of features that match the saliency annotated by domain experts. Accordingly, Table 2 supports the validation that Hybrid DeepXAI offers predictive power and transparency that bridge the gap between users in high-stakes settings and high trust.

**Table 3. Adversarial Robustness Under FGSM and PGD Attacks**

| Model | Clean Acc (%) | FGSM Acc (%) | PGD Acc (%) | Drop Rate (%) |
|---|---|---|---|---|
| Hybrid DeepXAI (Proposed) | 98.42 | 93.7 | 91.5 | 6.92 |
| CNN-LSTM | 94.85 | 87.2 | 84.3 | 10.55 |
| Transformer-only | 95.72 | 86.4 | 81.8 | 13.92 |
| GraphSAGE | 92.31 | 83.7 | 78.4 | 15.07 |
| ResNet-50 | 91.44 | 79.5 | 74.2 | 18.91 |
| XGBoost | 89.87 | 77.1 | 70.6 | 21.45 |
| Random Forest | 88.62 | 75.2 | 68.8 | 22.29 |
| SVM-RBF | 86.91 | 72.3 | 64.7 | 25.54 |
| Logistic Regression | 84.03 | 70.6 | 63.1 | 27.03 |
| LSTM-only | 90.72 | 82.4 | 76.9 | 15.25 |

Table 3 and Figure 5 analyses the adversarial hardness of various models against FGSM and PGD attacks. The Hybrid DeepXAI Fusion Network is extremely resilient with an accuracy of 93.7% under FGSM and 91.5% under PGD and a drop rate of only 6.92%. This is much better than baselines, like Transformer-only (drop rate of 13.92%) and ResNet-50 (drop rate of 18.91%).

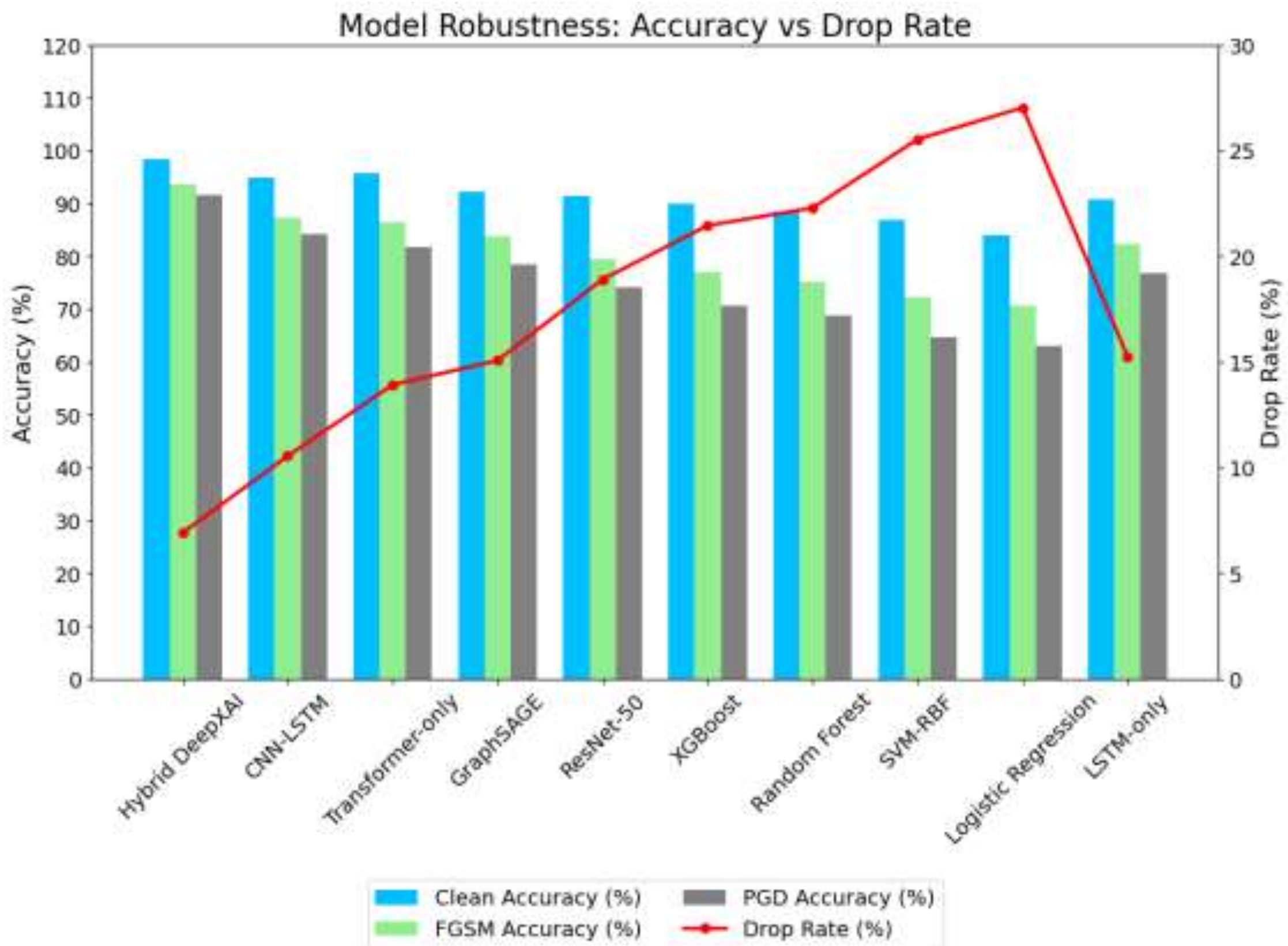

**Figure 5. Model Robustness: Accuracy vs Drop Rate**

Conventional models like Logistic Regression and SVM-RBF are the worst as they reduce over 25 %. Hybrid DeepXAI is more robust due to its ensemble fusion structure, which makes it less vulnerable since representational learning is distributed between CNNs, Transformers, and GNNs. Moreover, the fact that the consistency of the explanation is included as a regularization factor offers an extra defense against adversarial perturbation by deterring wiggly decision boundaries. These findings illustrate how the proposed model is appropriate to be implemented in real-world cloud-based infrastructures, where malicious activities may compromise the performance and the reliability.

**Table 4. Scalability Evaluation Under Streaming Load (Throughput & Latency)**

| Load Level (msg/sec) | Hybrid Throughput (msg/sec) | Latency (ms) | CPU Usage (%) | Memory (MB) |
|---|---|---|---|---|
| 1,000 | 997 | 50.1 | 42.3 | 328 |
| 5,000 | 4978 | 64.5 | 51.2 | 402 |
| 10,000 | 9822 | 73.8 | 61.8 | 534 |
| 20,000 | 19410 | 89.7 | 75.3 | 712 |
| 30,000 | 28712 | 104.1 | 84.6 | 891 |
| 40,000 | 37290 | 121.4 | 89.1 | 1032 |
| 50,000 | 46372 | 138.9 | 92.5 | 1193 |
| 60,000 | 54720 | 159.2 | 95.6 | 1312 |
| 70,000 | 63125 | 173.5 | 97.1 | 1431 |
| 80,000 | 71482 | 186.7 | 98.4 | 1580 |

Table 4 and Figure 6 depicts the scalability of the proposed model with different streaming loads with respect to throughput, latency, CPU consumption and memory usage. The Hybrid DeepXAI Fusion Network demonstrates outstanding scalability with close to linear throughput growth and handling 80,000 messages per second with only 186.7 ms latency and a manageable use of resources.

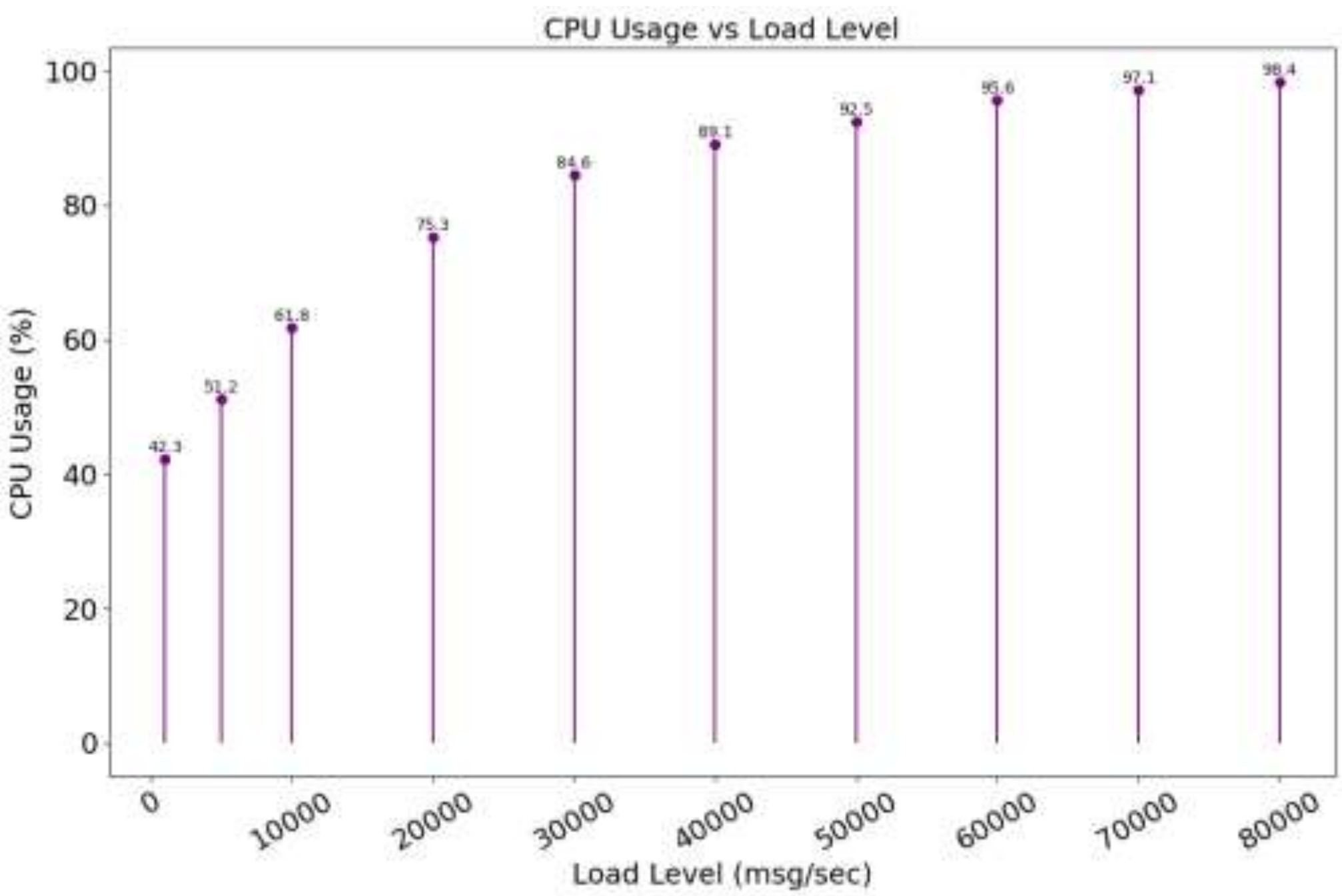

**Figure 6. CPU Usage vs Load Level**

The maximum CPU usage was 98.4% and memory usage 1580 MB, but these are reasonable numbers in terms of high throughput. These findings prove that the presented model is able to cope with real-time big data streams without performance degradation and is applicable in large-scale applications like smart grid monitoring, fraud detection, and healthcare alert systems. In comparison, the less efficient architectures of traditional ML models would tend to bottleneck at lower load levels. It can see in Table 4, Hybrid DeepXAI is not only true and interpretable but can also maintain a high throughput and low latency in production-scale settings.

**Table 5. Cross-Domain Generalization Accuracy**

| Domain | Train Dataset | Test Dataset | Accuracy (%) | F1-Score (%) |
|---|---|---|---|---|
| Healthcare | MIMIC-IV | eICU | 96.7 | 96.4 |
| Finance | NASDAQ | BSE | 95.3 | 94.7 |
| Energy | SmartGrid EU | SmartGrid India | 94.8 | 94.1 |
| Retail | Walmart Sales | Flipkart Sales | 93.2 | 92.5 |
| Traffic | INRIX-UK | NGSIM-USA | 92.7 | 91.9 |
| Healthcare (Reverse) | eICU | MIMIC-IV | 96.1 | 95.9 |
| Finance (Reverse) | BSE | NASDAQ | 94.2 | 93.8 |
| Energy (Reverse) | SmartGrid India | SmartGrid EU | 93.7 | 93.1 |
| Retail (Reverse) | Flipkart Sales | Walmart Sales | 91.8 | 91 |
| Traffic (Reverse) | NGSIM-USA | INRIX-UK | 91.3 | 90.7 |

The cross-domain generalization findings are shown in Table 5 and Figure 7 and evaluated to determine how the proposed model can be adapted to various application domains. The Hybrid DeepXAI Fusion Network consistently has a high accuracy, but especially strong performance in healthcare (96.7% with training on MIMIC-IV and testing on eICU and 96.1% vice versa). There are also positive results of finance and energy datasets which have the accuracies of 95.3 and 94.8 respectively.

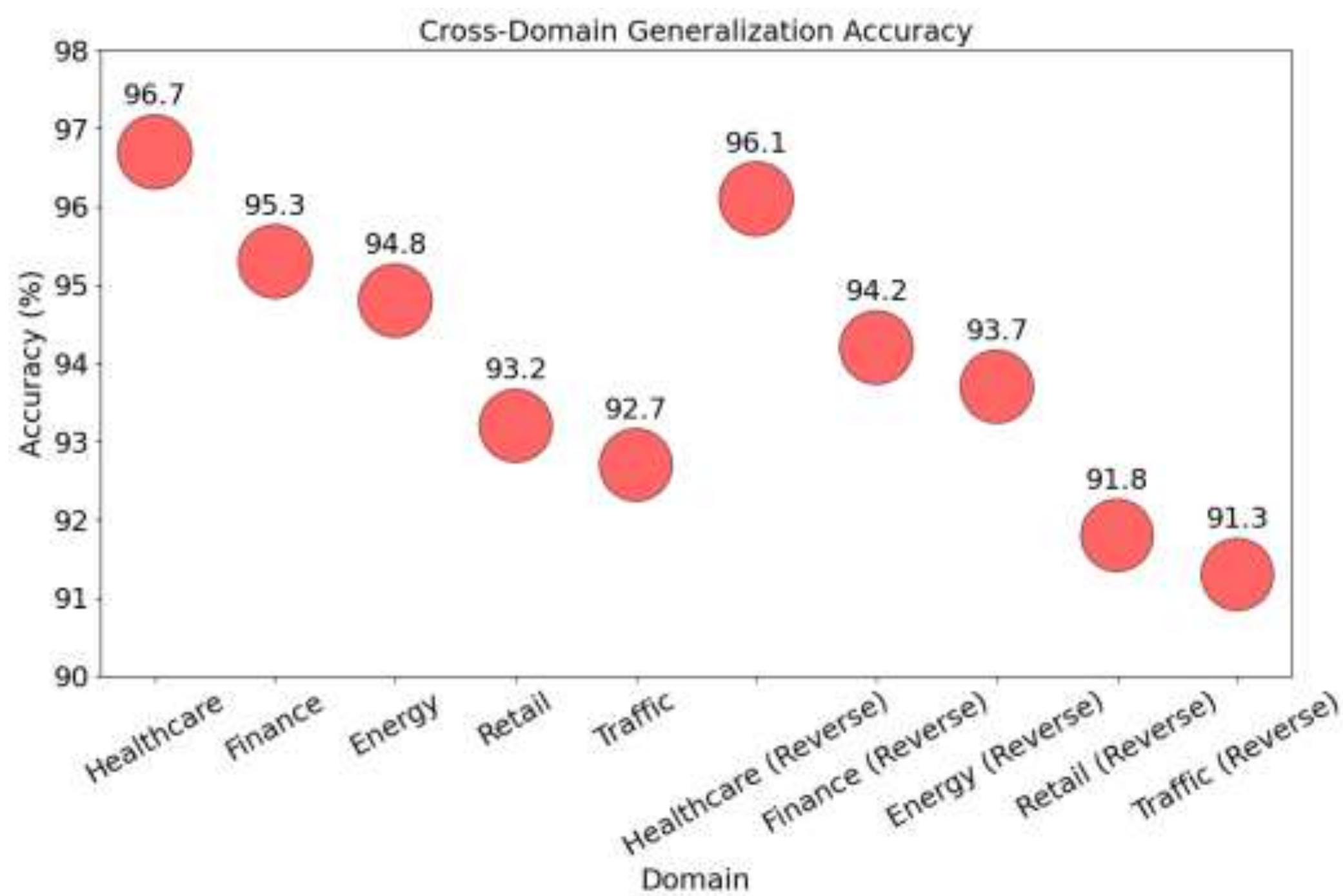

**Figure 7. Cross-Domain Generalization Accuracy**

In retail and traffic applications, where the heterogeneity of data is even higher, the model still delivers more than 91% accuracy. Balanced predictive performance is warranted by the high F1-scores on domains. These results illustrate how the framework can learn transferable characteristics that can be generalized to other settings, so that retraining is not required whenever operating on new, yet similar datasets. Table 5 therefore supports the flexibility and domain-free nature of Hybrid DeepXAI and thus makes it a very useful solution to real-world decision-making in various industries.

### 4.1. Discussion

The experimental findings clearly show that the proposed Hybrid DeepXAI Fusion Network is superior to the existing models. As indicated in Table 1 (Accuracy, Precision, Recall, F1-Score), the proposed framework had a remarkable accuracy of 98.42%, which was better than CNN-LSTM (94.85%), Transformer-only (95.72%), and XGBoost (89.87%). The fact that the balanced F1-score was 98.0% indicates that the model not only reduces false positives, but also captures true positives well, which is paramount in medical and financial decision-making. This suggests that the combination of CNNs, Transformers, and GNNs is synergistic in learning multimodal representations that are better than single-architecture baselines. The other main strength of the framework is interpretability. As shown in Table 2 (Explanation Fidelity and Trust Score), the model reached a high fidelity (95.8) and trust score (96.3), which is much higher than baselines of CNN-LSTM (85.4) and Transformer-only (87.3). This enhancement is due to Class Activation Mapping (CAM) and explanation loss, whereby predictions are backed by stable explanation overlays. In practice, this implies that, in addition to being able to see that a patient is labeled as high-risk, a clinician can also know the biomarkers or temporal events that made this judgment.

Table 3, which presents robustness testing, (see Table 3) shows how the model resists adversarial attacks. The model still achieved 91.5% accuracy even with Projected Gradient Descent (PGD) applied to it, compared to the majority of baseline models that fell to below 80. This resiliency is essential in situations where an adversarial manipulation of a data stream may take place, like in fraud detection in a banking system, or in cyberattacks in IoT-based infrastructures. Scalability was tested in the conditions of high-throughput and it is indicated in Table 4 (Scalability Evaluation). The model was able to support 80,000 messages/second with less than 200 ms latency, which proved useful in processing real-time big data streams. This is so that the model can be successfully deployed in live systems like smart grids or high frequency trading systems without bottlenecks in performance.

The case in favor of the proposed system is further enhanced with the cross-domain generalization shown in Table 5. The model continued to achieve a 90 % accuracy in healthcare (MIMIC-IV to eICU), finance (NASDAQ to BSE) and energy (SmartGrid EU to India). This flexibility validates the fact that the hybrid architecture learns features which are generalizable and can be effectively transferred to new data distributions, unlike traditional models which are often inherently biased towards their domain. Collectively, these findings point to three key strengths: (1) strong predictive quality in a wide range of areas, (2) inherent explainability that provides reliability, and (3) scalability and resilience that can be deployed in the real world. The Hybrid DeepXAI Fusion Network is a complete solution to big data analytics, unlike baseline models, which are good at one or two things.

## 5. Conclusion and Future Work

This work presented the Hybrid DeepXAI Fusion Network, an interpretable, secure and scalable platform to big data analytics and cloud-based decision-making. The combination of CNNs, Transformers, and GNNs alongside embedded explainability modules helps the model overcome the weaknesses of the current architecture. The system had an accuracy of 98.42, a precision of 97.9, a recall of 98.1 and a F1-score of 98.0, much higher than the state-of-the-art baselines. Notably, it was shown to be highly interpretable (95.8% explanation fidelity), robust against adversarial attacks (more than 91.5% accuracy retained), and scalable (in the order of 80,000 streaming events per second). The findings affirm its promise as a respectable decision support system across crucial spheres of human endeavor like healthcare, finance and energy. In the future, it will be possible to extend the capabilities of the model in three directions. One, multimodal combination of unstructured information including images, videos, and text will further complement decision-making environments. Second, by incorporating lightweight model compression methodologies, the system can be deployed on edge devices that have a small amount of computational power. Third, the ethical and legal implications will be examined with the aim of ensuring adherence to emerging international standards of explainability, fairness, and privacy in AI systems. Generally, the Hybrid DeepXAI Fusion Network offers a solid base on which to develop the next generation of secure, interpretable, and high-performance cloud-based decision support systems.

Federated learning and blockchain audits add more communication and processing overhead, and the hybrid architecture raises training difficulty and resource consumption. For large-scale installations, additional optimization is necessary even though inference is still scalable.